\documentclass[10pt,prc,aps,nofootinbib,showkeys,showpacs,twocolumn,superscriptaddress]{revtex4-1}
\usepackage{graphicx}
\usepackage{epsfig}
\usepackage{mathtools}
\usepackage[dvipdfm,colorlinks=true,urlcolor=blue,citecolor=blue,breaklinks=true]{hyperref}

\usepackage{color} 

\begin{document}
\title{On the formulation of functional theory for pairing with particle number restoration}
\title{Beyond mean-field calculation for pairing correlation}

\author{Guillaume Hupin} \email{hupin1@llnl.gov}
\affiliation{Grand Acc\'el\'erateur National d'Ions Lourds (GANIL), CEA/DSM-CNRS/IN2P3, Bvd Henri Becquerel, 14076 Caen, France}
\affiliation{Lawrence Livermore National Laboratory, P. O. Box 808, L-414, Livermore, California 94551, USA}
\author{Denis Lacroix} \email{lacroix@ganil.fr}
\affiliation{Grand Acc\'el\'erateur National d'Ions Lourds (GANIL), CEA/DSM-CNRS/IN2P3, Bvd Henri Becquerel, 14076 Caen, France}

\begin{abstract}
The recently proposed Symmetry-Conserving Energy Density Functional approach [G. Hupin, D. Lacroix and M. Bender, \href{http://dx.doi.org/10.1103/PhysRevC.84.014309}{Phys. Rev. C84, 014309 (2011)}] is applied to perform Variation After Projection onto good particle number using Skyrme interaction, including density dependent 
terms.  \\
We present a systematic study of the Kr and Sn isotopic chains. This approach leads to non-zero pairing in magic nuclei and 
a global enhancement of the pairing gap compared to the original theory that breaks the particle number symmetry. 
The need to consistently readjust the pairing effective interaction strength is discussed. 
\end{abstract}

\date{28 November 2011}


\maketitle

\section{Introduction}
\label{sec:intro}

The nuclear Energy Density Functional (EDF) is a versatile approach \cite{Rin80, Ben03} 
that allows one to describe a variety of phenomena in nuclei ranging from nuclear structure effects, to nuclear dynamics, and 
thermodynamics. One specificity of nuclear energy functional approaches is that the densities 
used in the energy might not respect some of the properties related to the symmetry of the underlying bare many-body Hamiltonian \cite{Rei84,Naz92}.
This is achieved by introducing a reference Slater (or quasi-particle) state from which the one-body normal density
(and eventually the anomalous density) is constructed to express the energy. These densities are generally localized in space 
and therefore do not correspond to a translationally invariant system. Symmetry breaking is often extended to states that are neither an eigenstate of the particle 
number operator (then breaking the U(1) symmetry) nor of the total angular momentum operator. 

Symmetry-Breaking (SB)-EDF is a powerful technique to describe some aspects of nuclei like 
the onset of pairing and/or deformation. First however, restoration of broken symmetries is necessary to compare with experiment, where eigenstates with good quantum numbers are probed. Second, the restoration of symmetries 
and more, in general, the use of configuration mixing techniques is a way to grasp some additional correlations associated 
with quantum fluctuations in a collective space\cite{Egi82a,Egi82b}. Ultimately, the state of the art of EDF is to perform a configuration mixing to describe the coexistence of different intrinsic configurations like shapes, excited states, electromagnetic and nuclear transitions.

This technique of symmetry breaking followed by symmetry restoration has been recently shown to lead to spurious 
contributions to the energy and must be applied with caution \cite{Ang01,Dob07}. Overall, the very notion of symmetry-breaking in a functional 
approach needs to be clarified \cite{Dug10}. 
For a detailed discussion, we refer the interested reader to the recent works of refs. \cite{Ang01,Dob07,Lac09,Ben09,Dug09,Rob10}.  Facing these difficulties, at present, three 
strategies have been proposed to perform
well converged configuration mixing calculations within EDF: (i) Derive the energy functional starting from 
a true Hamiltonian and completely incorporate the Pauli principle \cite{Ang01}. (ii) Identify and remove spurious contributions from the energy functional \cite{Lac09,Ben09,Dug09}. This could be performed for some specific functionals, by comparing with the 
Hamiltonian case. (iii) Consistently extend the energy 
functional used in the SB case to a functional of the densities of the state with the symmetries restored. The latter strategy is the Symmetry-Conserving
(SC)-EDF approach proposed in ref. \cite{Hup11}. 

The strategies (i) and (ii) prevent us from using density dependent interactions with non-integer powers of the density, and strongly reduce 
the ability to tailor the density functional. Note that strategy (i) is nowadays used  with the Gogny force \cite{Rod07, Rod10,Lop11} taking 
specific care of the density dependent term. Recent applications of strategy (ii) have shown that this approach becomes rather 
involved when several symmetries are restored simultaneously\cite{Ben12}. While currently formulated 
only for the particle number restoration case (see also discussion in ref. \cite{Hup12}), the strategy (iii) can be used for any functional 
form such as those used in the SB case starting from the Gogny or Skyrme like interaction, while having a different interaction 
in the pairing channel. In addition, it is not required to strictly 
enforce the anti-symmetrization and some useful numerical approximations like the Slater approximation for the Coulomb exchange can be still used. 

In this article, the work presented in ref. \cite{Hup11} is extended to perform Variation After Projection (VAP), 
enforcing good particle number. We show that the SC-EDF used with the up-to-date functionals based on Skyrme interaction 
can be competitive to describe pairing in nuclei.   

\section{The Symmetry-Conserving EDF approach}

Starting, from a quasi-particle state $| \Phi_0 \rangle$, most currently used Symmetry-Breaking EDF based 
on the Skyrme\cite{Sky59} or Gogny\cite{Gog70} forces can be written as 
\begin{align}
{\cal E}_{\rm SB}[\Phi_0] =&  \sum_{i} t_{ii} ~\rho_{ii}+ \frac{1}{2} \sum_{i,j} \overline{v}_{ijij}^{\rho \rho}  ~\rho_{ii}\rho_{jj}  \nonumber \\
&+ \frac{1}{4} 
 \sum_{i,j} \overline{v}_{i\bar\imath j\bar\jmath }^{\kappa \kappa}~ \kappa_{i \bar\imath }^* \kappa_{j \bar\jmath } \; ,
\label{eq:denssr}
\end{align}
where $\overline{v}^{\rho \rho}$ and $\overline{v}^{\kappa \kappa}$ denote the effective vertices 
in the particle-hole and particle-particle channels. Here $\rho$ and $\kappa$ denote the normal and anomalous 
densities expressed in the canonical basis.

In the Hamiltonian case, the U(1) symmetry restoration can be performed by considering 
the component of the quasi-particle state with specific particle number. Introducing the particle 
number projector $P^N$, a new state $| \Psi_N \rangle$ defined through:
\begin{eqnarray}
| \Psi_N \rangle = P^N | \Phi_0 \rangle \, .
\label{eq:projstate}
\end{eqnarray}   
While there is no ambiguity when a Hamiltonian is used, the main challenge within EDF is to 
properly extend (\ref{eq:denssr}) to account for particle number conservation.  

This problem has been carefully analyzed in ref. \cite{Hup11} leading to a generalization 
of the energy density functional given by:
\begin{align}
\mathcal{E}_{\rm SC}[\hat \rho^{N}, \hat R^{N} ]  =& \sum_{i} t_{ii}  ~ \rho^{N}_{ii} + \frac{1}{2} \,  \sum_{i\ne j, j\neq \bar\imath } \bar v_{ijij}^{\rho \rho}  ~ R^{N}_{jiji} \nonumber \\ 
&+  \frac{1}{2} \, \sum_{i} \left(\bar v_{iiii}^{\rho \rho} + \bar v_{i\bar\imath  i \bar\imath }^{\rho \rho}\right)  ~ (\rho^{N}_{ii})^2 \nonumber \\
&+ \frac{1}{4} \, \sum_{i \neq j, i  \neq \bar \jmath } \bar v_{i\bar\imath j\bar\jmath }^{\kappa \kappa}  ~  R^{N}_{ \bar\jmath j\bar\imath i} 
\nonumber \\ 
&+  \frac{1}{2} \, \sum_{i} \bar v_{i\bar\imath i\bar\imath }^{\kappa \kappa} ~ \rho^{N}_{ii}\, (1 - \rho^{N}_{ii}) \; ,
\label{eq:scedf}
\end{align}
where $\rho^{N}$ and $R^{N}$ denote respectively the one and two body matrices of the projected state, Eq. (\ref{eq:projstate}).  
Note that, within SC-EDF, it is further postulated that any dependence 
of the effective vertices in terms of the SB density should be replaced by an equivalent dependence of 
the projected density. Doing so, the energy becomes a functional of the projected state degrees of freedom (DOF) only.
In former applications of SC-EDF, the expression of the energy has been used in the Projection After Variation (PAV) 
scheme, showing the absence of any pathologies previously observed, even if density-dependent interaction are used in the functional.

Here, the SC-EDF is applied to perform VAP. In this case, the energy should be minimized 
with respect to all possible variations of the projected state DOF, i.e. 
\begin{eqnarray}
\delta \mathcal{E}_{\rm SC}[\hat \rho^{N}, \hat R^{N} ]  &=& 0.
\end{eqnarray}

In the following we will consider the specific case where the state $| \Phi_0 \rangle$
is written in a BCS form as
\begin{eqnarray}
| \Phi_0 \rangle &=& \prod_{i>0} \left( u_i + v_i ~ a^\dagger_i a^\dagger_{\bar\imath } \right) | 0 \rangle \, ,
\label{eq:bcsstate}
\end{eqnarray}
with $u^2_i + v^2_i = 1$. Accordingly, variation of the projected state DOF can be recast into variations of the single-particle 
state components $\phi_i({\bf r})$ associated to the creation operator $ a^\dagger_i$ and variations of the quantity $v^2_i$ corresponding to the SB occupancy of orbital $i$. We then end 
up with a set of coupled equation to be solved self-consistently:
\begin{eqnarray}
\frac{\delta \mathcal{E}_{\rm SC}}{\partial \phi_i^\star({\bf r})} & = & 0,  ~~~~~~~
\frac{\delta \mathcal{E}_{\rm SC}}{\partial v^2_i}  =  0 \; .
\end{eqnarray} 
This procedure is the same as the one generally 
used in the Hamiltonian case in PNP-VAP \cite{Die64,She00,She01}. The eigenvalue equations of the self-consistent problem are recalled and explained in Appendix \ref{appA}.
It is worth mentioning that we took advantage of the analytic expressions of the densities $R^N$\cite{Lac10}. This step is crucial to reduce the computational burden of the calculation. 

The Euler-Lagrange equations associated with the minimization of the energy each yield 
a set of eigenvalues and non-linear equations that are rather involved numerically. Taking advantage
of the EDF flexibility without breaking the consistency requirement of the approach \cite{Hup11}, the minimization can be greatly 
simplified numerically by making the assumption in Eq. (\ref{eq:scedf}) that,
\begin{eqnarray}
 R^{N}_{jiji} &\simeq & \rho_{ii}^N \rho_{jj}^N \; . \label{eq:prescription}
\end{eqnarray}
This approximation is used  in the following.

\section{Applications}

The {\rm EV8} code of Bonche, Flocard and Heenen \cite{Bon05} has been updated to allow minimization 
of the  functional \ref{eq:scedf} using the approximation in Eq. (\ref{eq:prescription}). 
The numerical method consists in solving the mean field problem by an imaginary time step method \cite{Dav80} and the optimization of the occupation probabilities by a sequential quadratic programming.
In the following, the SC-EDF method is used with the SLy4 interaction in the mean-field 
channel \cite{Cha98} while the effective pairing interaction considered \cite{Cha76} is
\begin{eqnarray}
v^{\kappa \kappa}(\mathbf{r},\mathbf{r}') &=&\frac{V_0}{2}~ (1 - P_{\sigma}) ~ \left( 1- \left( \frac{{\rho}\left( \mathbf{R} \right)}{\rho_0}\right)^\alpha \right)~ \delta(\mathbf{r}-\mathbf{r}') \, ,\nonumber \\ \label{eq:pairdd}
\end{eqnarray}
with $\mathbf{R} =(\mathbf{r}+\mathbf{r}')/2 $. $V_0= 1250$ MeV is the pairing constant, $\alpha = 1$ and $\rho_0 =0.16$ fm$^{-3}$ is the saturation density. In addition, 
to avoid the ultra-violet divergence that appears with contact interaction, a cut-off factor \cite{Bon85} with an energy interval of $5$ MeV is used to select states around the Fermi energy. These values 
have been typically used to reproduce neutron and proton separation energies \cite{Rig99} and, in the standard terminology, 
correspond to a surface pairing.

\begin{figure}[ht]
\includegraphics[width=8.cm]{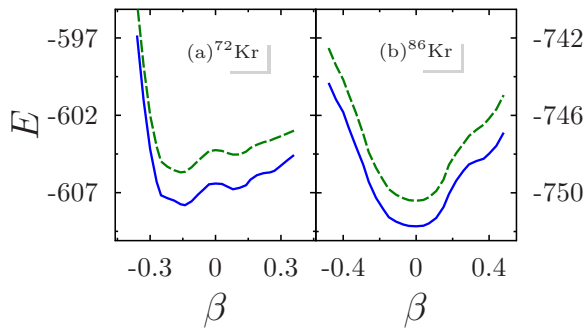}
\caption{\label{fig:krpes}
(Color online) Evolution of the energy obtained using the VAP calculation (blue solid line) for the $^{72}$Kr (a) and $^{86}$Kr
(b) as a function of deformation. In each case, the BCS result obtained with the original EV8 code is shown with a green dashed curve.
}
\end{figure}

\begin{figure}[ht]
\includegraphics[width=8.cm]{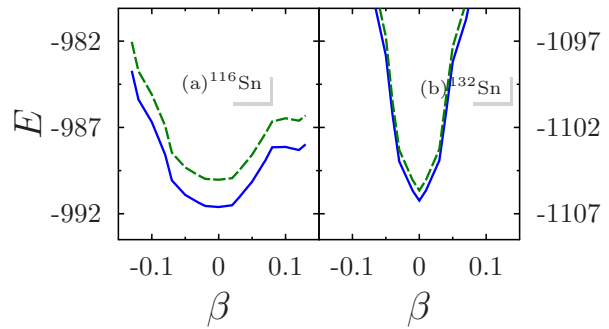}
\caption{\label{fig:snpes}
(Color online) Same as figure \ref{fig:krpes} for the $^{116}$Sn (a) and $^{132}$Sn (b).
}
\end{figure}

In this work, SC-EDF calculations are systematically performed for the Kr and Sn isotopic chains. 
In the latter case, the proton number is magic while in the former case it is not. 
As an illustration of the results, the evolution of the energy as a function of the deformation 
obtained with the SC functional (blue solid line) is compared to the original BCS result (green dashed line) 
for $^{72}$Kr and $^{86}$Kr respectively in panel (a) and (b) 
of figure \ref{fig:krpes}. Similar curves are shown in figure \ref{fig:snpes} for $^{116}$Sn and $^{132}$Sn. These 
nuclei have been selected because they are representative of the different types of situations: mid-shell nucleus ($^{72}$Kr), 
simply ($^{86}$Kr and $^{116}$Sn) or doubly magic nucleus ($^{132}$Sn).
The results have been obtained by
adding a quadrupole constraint in the minimization while the deformation parameter is defined by
\begin{eqnarray}
 \beta &=& \sqrt{\frac{5}{16 \pi}} ~ \frac{4 \pi}{3R^2 A} ~ \langle Q_{20} \rangle \, , \label{eq:defor}
\end{eqnarray}
where $\langle Q_{20} \rangle$ is the quadrupole deformation and $R=1.2~ A^{1/3}$ is the nuclear radius. 
It can be seen in figures \ref{fig:krpes} and \ref{fig:snpes}, that the potential energy curves 
obtained with the SC-EDF are smooth and, as already discussed in ref. \cite{Hup11}, we do not see any of the 
pathologies observed in other approaches with density dependent interactions \cite{Dob07,Dug09}.

Figures \ref{fig:krpes} and \ref{fig:snpes} illustrate that the energy potential curves of the SC-EDF functional with respect to the quadrupole deformation are shifted from BCS. There are no changes in the shape of these curves, both BCS and shifted SC functional can be almost superimposed. The energy gain, illustrated by the shift, is between $1$ and $2$ MeV for mid-shell and simply magic nuclei while the doubly magic nuclei $^{132}$Sn gains more than $0.5$ MeV. This increase in correlation energy comes from the improved treatment of the pairing correlations from the projection formalism that has been used to tailor the functional dependances of the energy.  %\textcolor{blue}{The pairing energy can also be plotted here and then compared to the one shown by Rodriguez. He studied the Cr and saw a shift of the PES due to the VAP.}

As seen from figures \ref{fig:krpes} and \ref{fig:snpes},  the full SC functional induces rather small differences on the total energy 
compared to the original BCS case. It should be mentioned, however, that the pairing energy 
is always enhanced when the symmetry is conserved, especially around shell closures, as expected.  
Indeed, when the pairing is treated within BCS or HFB, there is a sudden disappearance of correlations in the weak pairing 
regime. This is the well-known BCS threshold anomaly. 
A measure of the pairing strength is provided by the mean gap \cite{Dob84}:
\begin{eqnarray}
\overline{\Delta}_{n/p} &=& \frac{E^{n/p}_{\rm pairing}}{\sum_i ~ \sqrt{\rho_{ii} (1 - \rho_{ii})}}\, , \label{eq:meangap}
\end{eqnarray}
where $\rho_{ii}$ are the occupation probabilities of a given theory and $E^{n/p}_{\rm pairing}$ its neutron/proton pairing energy. 
In the SC-EDF, these energies are calculated as the sum of the last two terms in Eq. (\ref{eq:scedf}).
This observable has the advantages of (i) correlating with the pairing gap in the limit of a constant pairing interaction and (ii) probing both the pairing energy and the trend of the occupation probabilities, such as the fragmentation of occupation numbers around the Fermi surface.

\begin{figure}[ht]
\includegraphics[width=8.cm]{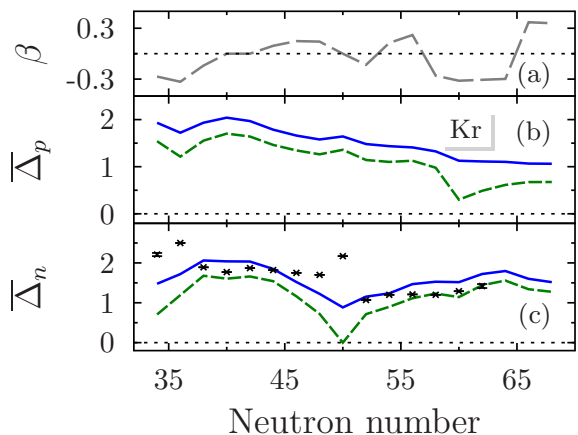}
\caption{\label{Mgapg1-Kr}
(Color online) 
Value of the deformation parameter $\beta$ (Eq. (\ref{eq:defor})) at the minimum (a), 
average proton (b) and neutron (c) mean gaps defined by Eq. (\ref{eq:meangap}) are shown as a function of the neutron number along the Kr isotopic chain obtained from BCS (green dashed lines) and the SC-EDF (blue solid lines). The deformation parameter at the minimum is the same for both BCS and SC-EDF (gray long dashed line). The calculations are performed with a SLy4 effective interaction that includes a non-integer density dependence and a density dependent pairing interaction(Eq. (\ref{eq:pairdd})). The minimization is performed including the quadrupole degree of freedom. In the neutron case, the experimental gaps (black crosses) and their error bars\cite{nndc} obtained with the three points formula(see \cite{Dug01,Dob84}) are also presented.
}
\end{figure}
\begin{figure}[ht]
\includegraphics[width=8.cm]{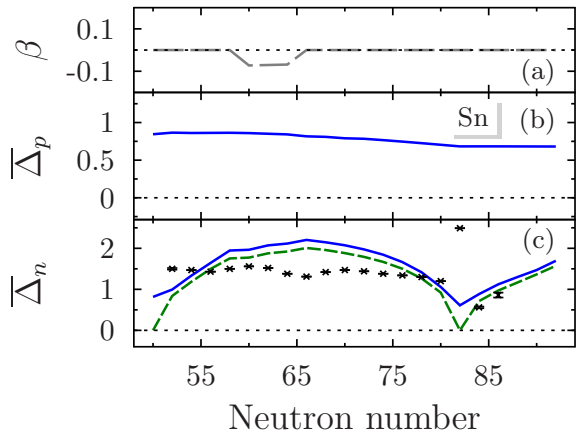}
\caption{\label{Mgapg1-Sn}
(Color online)
Same as figure \ref{Mgapg1-Kr} for the Sn isotopic chain. The proton mean gap (b) of BCS identifies with zero along the isotopic line.}
\end{figure}

In figures \ref{Mgapg1-Kr} and \ref{Mgapg1-Sn}, the deformation 
parameter $\beta$ (Eq. (\ref{eq:defor})) at the minimum  of the energy (\ref{Mgapg1-Kr}a and \ref{Mgapg1-Sn}a), and 
the average proton (\ref{Mgapg1-Kr}b and \ref{Mgapg1-Sn}b) and neutron (\ref{Mgapg1-Kr}c and \ref{Mgapg1-Sn}c) gaps are shown as a function of the neutron number along the Kr and Sn isotopic chains, respectively. 
The BCS (green dashed lines) and the SC-EDF (blue solid lines) results are compared. Note that consistently with the observations from figures \ref{fig:krpes} and \ref{fig:snpes}, the deformation parameter at the minimum of the energy (long dashed line) is the same for both BCS and SC-EDF (hence is plotted once).
In this figure, BCS exhibits strong variations of the gap near the $N=50$ shell closure. This is a fingerprint of the abrupt 
disappearance of pairing in this formalism close to magicity. It is also worth to keep in mind that the evolution of deformation 
as $N$ increases might also induce local fluctuations. This is the case for $N > 56$ in the Kr chain as we can see from the evolution of the deformation (\ref{Mgapg1-Kr}a) reflected by variations in panels (\ref{Mgapg1-Kr}b) and (\ref{Mgapg1-Kr}c).

In the SC-EDF case, it is observed that the pairing gap is systematically enhanced compared to the BCS results. 
This enhancement is increased at the shell closure. For instance, in the proton gap of Sn isotopes, a mean gap
of $\sim 0.7$ MeV is obtained (see figure \ref{Mgapg1-Sn}) compared to zero MeV at the BCS level. 
In the Kr isotopic chain, both BCS and SC-EDF lead to deformed nuclei with the same deformation parameter. The increase of pairing correlations is only due to a better treatment of quantum fluctuations in gauge space by the SC method. 
It is then seen that the increase at the shell closure ($N=50$) is further enhanced to $\sim 1$ MeV, while it is of the order of $0.3-0.5$ MeV in the mid-shell. Altogether, the pairing gap obtained within the VAP approach is much smoother than the 
BCS pairing gap and more consistent with experimental observations.

\begin{figure}[ht]
\includegraphics{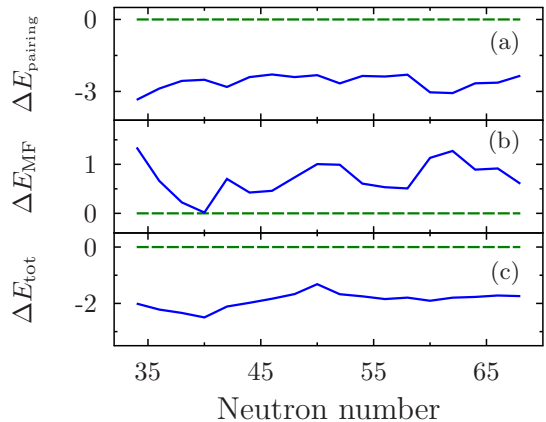}
\caption{\label{fig:dekr}
(Color online) Evolution of the quantities $\Delta E_{\rm pairing}$ (a), $\Delta E_{\rm MF}$ (b) 
and $\Delta E_{\rm tot} $ (c)  along the Kr isotopic chain. The horizontal dashed line corresponds to the case where 
BCS and SC-EDF would be identical.
}
\end{figure}

\begin{figure}[ht]
\includegraphics{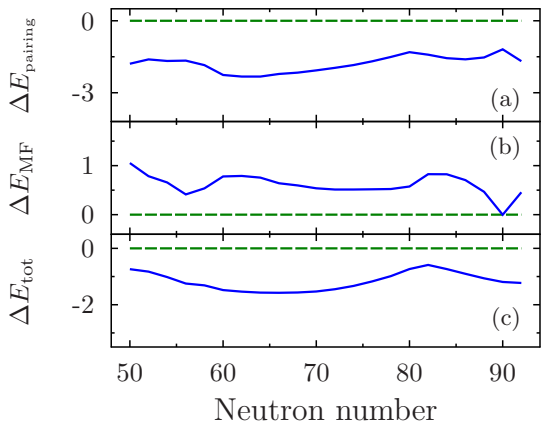}
\caption{\label{fig:desn}
(Color online) Same as figure \ref{fig:dekr} for the Sn isotopic chain.}
\end{figure}

It is important to note that the increase of the pairing gap is not fully reflected in the lowering of the
ground state binding energy. Indeed, the SC-EDF is a fully self-consistent approach and, when the enhanced pairing is built up 
in the minimization, the mean-field reorganizes. Generally, it is observed that the mean-field energy, denoted by $E_{\rm MF}$ 
and defined as the total 
energy minus the pairing energy, increases slightly and partially compensates for the effect of the pairing. In figures \ref{fig:dekr} 
and \ref{fig:desn}, the three quantities 
\begin{eqnarray}
\Delta E_{\rm pairing} & = & E^{\rm VAP}_{\rm pairing} -E^{\rm BCS}_{\rm pairing} \, ,\nonumber  \\
\Delta E_{\rm MF}  & = & E^{\rm VAP}_{\rm MF} -E^{\rm BCS}_{\rm MF} \, ,\nonumber  \\
\Delta E_{\rm tot}  & = & E^{\rm VAP}_{\rm tot} -E^{\rm BCS}_{\rm tot} \, ,\nonumber
\end{eqnarray} 
are displayed as a function of the neutron number respectively from panel (a) to (c) for the Kr and Sn isotopes. 
In these figure, we see that $\Delta E_{\rm pairing}$ (\ref{fig:dekr}a and \ref{fig:desn}a)  is always negative while $\Delta E_{\rm MF}$ (\ref{fig:dekr}b and \ref{fig:desn}b) is always positive 
and therefore, the net reduction of the total energy (c) is much less than the pairing correlation would suggest.
Altogether, the total energy is shifted. The transition from a sharp Fermi distribution around single or doubly magic nuclei with BCS to a 
fragmented Fermi surface with non-zero pairing within SC-EDF leads to a significant change in the mean-field energy, especially due to the contribution of single-particle levels above the Fermi energy. We can observe this effect in both figures \ref{fig:dekr} and \ref{fig:desn}. However, it is not possible to give more general trends because of the deformation and self-consistency of the theory.

\begin{figure}[ht]
\includegraphics[width=8.cm]{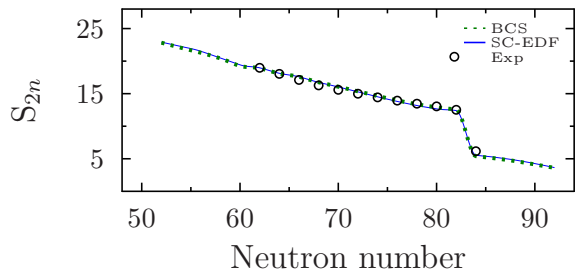}
\caption{\label{fig:s2n}
(Color online)  Comparaison of the two neutron separation energies S$_{2n}$ along the Sn isotopic chain between BCS (green dotted line) and SC-EDF (blue solid line) with the experimental values (black open circles).}
\end{figure}

In figure \ref{fig:s2n}, the two neutron separation energies S$_{2n}$
obtained in the BCS (green dashed line) and SC-EDF (blue solid line) are compared with experimental values (black open circles). This quantity is sometime used in the literature to adjust the 
pairing effective interaction parameters.  
Both BCS and VAP are consistent with experiment. In fact, the $S_{2n}$ are not affected by the variation after projection performed within SC-EDF.

The applications of the SC-EDF functional show that the bulk properties (figures \ref{fig:krpes}, \ref{fig:snpes} and \ref{fig:s2n}) of the underlying effective interaction are conserved while the total binding and pairing energies are shifted (figures \ref{fig:dekr} and \ref{fig:desn}). For all nuclei studied here, the SC-EDF functional predicts a non zero pairing energy and a fragmented Fermi surface. This is reflected by the non zero pairing gap (figures \ref{Mgapg1-Kr} and \ref{Mgapg1-Sn}) for all nuclei, including single and doubly magic ones where BCS leads to a Fermi distribution for the orbital occupancies. In the following, the evolution of these observations are investigated as a function of the refitting of the strength $V_0$ of the pairing interaction.

\section{Discussion of the pairing strength}

The pairing interaction used above is often adjusted to properly describe pairing gaps in EDF using BCS or HFB especially in the mid-shell \cite{Dob84,Dob95,Les06,Mar07}. As seen in the previous section, going beyond the  mean-field leads to an overestimation of the pairing energy in this region. Indeed, consistently with a density functional approach, one should {\it a priori} readjust the pairing strength when the functional changes. 
In this section, the results of VAP with an optimal value of the pairing strength are presented. 

%Since a global optimization of the SC-EDF  is not the aim of the present work, we focus on refitting the pairing interaction such as to reproduce 
%pairing gaps as well as the original BCS parametrization does. Also, the evolution of the potential energy versus deformation as a function of 
%the pairing regime is explored.

\begin{figure}[ht]
\includegraphics[width=8.cm]{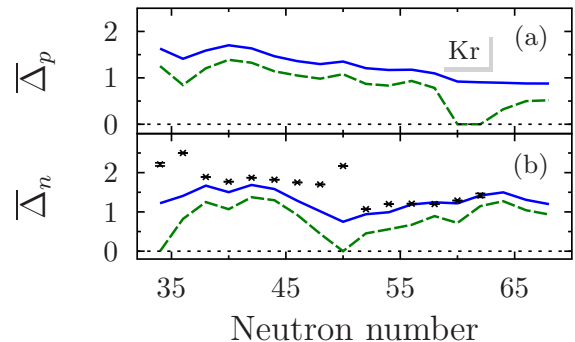}
\caption{\label{Mgapg0,8-Kr}
(Color online)
Same as figure \ref{Mgapg1-Kr} with a pairing strength of $V_0=1100$ MeV. The green dashed curve corresponds to the BCS result while the blue solid 
line corresponds to the SC-EDF case. 
}
\end{figure}
\begin{figure}[ht]
\includegraphics[width=8.cm]{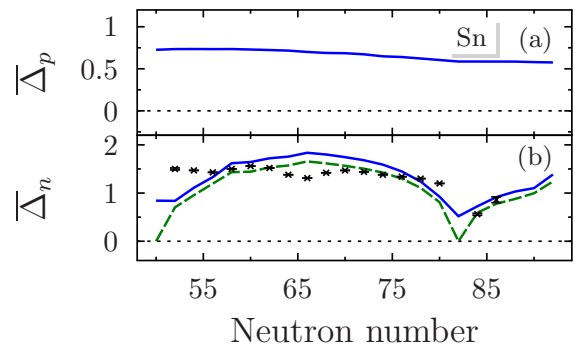}
\caption{\label{Mgapg0,8-Sn}
(Color online)
Same as figure \ref{Mgapg1-Sn} for the Sn isotopes.}
\end{figure} 

In figures \ref{Mgapg0,8-Kr} and \ref{Mgapg0,8-Sn},  results of the BCS (green dashed line) and SC-EDF (blue solid line) 
with a pairing strength $V_0=1100$ MeV are shown. This value of the strength has been chosen to properly describe Kr isotopes in the open-shell. 
By comparing these figures with figures \ref{Mgapg1-Kr} and \ref{Mgapg1-Sn}, it can 
be observed that SC-EDF with the reduced pairing interaction reproduces 
the original BCS result (with $V_0=1250$ MeV) in open shell nuclei.  

However, it should be mentioned that the refitting of pairing effective interaction solely when incorporating 
particle number projection is a too simplistic strategy 
to properly  describe both the pairing and the bulk properties in nuclei. This is illustrated in Table 
\ref{tab:refit}, where the binding and pairing energies obtained with BCS and the original pairing interaction 
are compared with the SC-EDF results and the reduced pairing strength. 
\begin{table}[hb]
\caption{Comparison between the binding and pairing energies of selected Kr isotopes 
%presented in figure \ref{KrDefo} 
predicted by BCS (with $V_0=1250$ MeV ) and SC-EDF (with $V_0=1100$ MeV). }.
\vspace{15pt}
\begin{tabular}{ c | c | c | c | c | c | c}
   \hspace{4pt}N\hspace{4pt}   &  \hspace{4pt}Z\hspace{4pt}  &  \hspace{4pt}BCS\hspace{4pt}  & BCS pairing    &  \hspace{4pt}VAP\hspace{4pt} & VAP pairing  & Exp\cite{nndc}   \\[0.15cm]
  \hline\noalign{\smallskip}
  \hline
  46 & 36 & -715.64 & -7.44 & -716.34 & -6.98 & -714.27 \\[0.08cm] 
  48 & 36 & -733.43 & -5.19 & -734.29 & -5.44 & -732.26\\[0.08cm] 
  50 & 36 & -750.41 & -4.99 & -750.91 & -5.49 & -749.23\\[0.08cm] 
  52 & 36 & -761.65 & -5.04 & -762.49 & -5.47 & -761.80\\[0.08cm] 
  54 & 36 & -772.3  & -5.9  & -773.18 & -5.63 & -773.21
\end{tabular}\label{tab:refit}
\end{table}
While pairing energies are similar, as could be anticipated from the fact that the gaps are similar, the binding energy 
deduced with the SC-EDF is often lower than the BCS and the original BCS is closer to the experimental binding energy.   
This stems from the fact that both the mean field and pairing channel have been consistently adjusted simultaneously 
at the BCS level. When performing VAP not only the pairing correlations are affected but also additional correlations build
up in the particle-hole channel, leading to a rather significant reorganization of the mean-field itself. This has been already 
clearly applied
 in figures \ref{fig:dekr}  and \ref{fig:desn}.  As a consequence, to improve the quality of theories 
that go beyond mean-field by restoring symmetries compared to those where symmetries are broken, in the near 
future, a complete readjustment of all components of the functional (mean-field and pairing)  should be considered at the VAP level.

\section{Conclusion}

The recently proposed Symmetry-Conserving EDF approach to incorporate the effect of particle number conservation is performed 
in the Variation After Projection (VAP) scheme. The VAP is applied using density dependent interaction both in the mean-field
and pairing channels. Such a density dependence, while impossible to use in configuration mixing calculations, does not lead to any difficulty in
the SC-EDF framework. 
Systematic study of the krypton and tin isotopic chains is made showing 
the increase of pairing energy when particle number conservation is taken into account self-consistently. In particular, the description of
correlations in the vicinity of closed shell nuclei is improved.  Indeed, as expected, the symmetry conserving theory predicts non vanishing pairing gaps around and at shell closures. The present study clearly shows that the incorporation of symmetry restoration leads to an enriched functional and that the parameters used to design the functional in the original symmetry breaking approach need to be 
consistently readjusted. Here, a first attempt is made to reduce the pairing strength in order to properly describe pairing gaps. We point out 
that, ultimately, coefficients of the functional in both mean-field and pairing channels should be simultaneously optimized  to really gain 
in predictive power of EDF approaches. 

%\bibliographystyle{apsrev4-1}
%\bibliography{VAP-Func}

\acknowledgements

This work in part performed under the auspices of the U.S. Department of Energy by Lawrence Livermore National Laboratory under Contract DE-AC52-07NA27344.

\appendix

\section{Euler-Lagrange equations}\label{appA}

In the case of the SC-EDF built from a quasi-particle vacuum  and a two-body delta interaction, the eigen-equations to be solved as a self-consistent mean-field problem read
\begin{align}
\frac{\partial \mathcal{E}_{\rm SC}}{\partial \phi_i^\star ({\bf r })} =& \Big( -  \frac{\hbar^2}{2 m} ~ { \Delta} + \sum_{j\ne( i, \bar \imath)} \frac{\partial ~ \bar v^{\rho\rho}_{ijij}}{\partial \phi_i^\star ({\bf  r })\partial \phi_i ({\bf  r })} ~ ~\frac{R_{jiji}^N}{\rho_{ii}^N}  \nonumber \\
&  +  \frac{\partial ~ \bar v^{\rho\rho}_{iiii}}{\partial \phi_i^\star({\bf r }) \partial \phi_i( {\bf r  })} ~ \rho_{ii}^N  - \varepsilon_i ~  \Big) ~~\rho_{ii}^N ~ \phi_i( {\bf r  })\; , \label{eq:valprop1}
\end{align}
where the contribution from the pairing part of the functional have been neglected as it is usually done, $\bar v^{\rho\rho}$ is a particle-hole contact interaction, $R^N_{jiji}$ is the projection of the one body density acting in the particle-hole channel and $\varepsilon_i$ is the Lagrange multiplier that enforces the normalization of the single-particle state $\phi_i$. It can be noted that in this form there is one potential for each orbitals due to the density dependence in the summation. The role of prescription Eq. (\ref{eq:prescription}) is to remove this dependency, hence recovering a single mean-field for all orbits.

\end{document}